\documentclass[11pt]{article}
\usepackage{graphicx}
\usepackage[colorlinks=true,
            breaklinks=true,
            urlcolor=blue,
            linkcolor=green,
            citecolor=green,
            bookmarks=true,
            pdftitle={Numerical Simulations of Breaking
               Waves-- Weak Spilling to Strong Plunging},
            pdfauthor={Kyle A. Brucker},
            pdfsubject={Proceedings of the 28th Naval Hydrodynamics
                         Symposium},
            pdfkeywords={Breaking Waves, Air Entrainment, Numerical
                         Simulations, High Performance Computing},
            bookmarksopen=false,
            pdfpagemode=None]{hyperref}
\pdfoutput=1
\begin{document}
\title{Numerical simulation of transom-stern waves} 
\author{Randall E. Hand$^1$, Miguel Valenciano$^1$, \\
Kevin George$^1$, Tom Biddlecome$^1$, Richard Walters$^1$, Mike Stephens$^1$,\\
Thomas T. O'Shea$^2$, Kyle A. Brucker$^2$, and Douglas G. Dommermuth$^2$\\ 
\vspace{4pt} \\
$^1$ Unclassified Data Analysis and Assessment Center, U.S. Army\\
Engineering Research and Development Center, MS 39180 \\
\vspace{2pt}\\
$^2$  Science Applications International Corporation\\
10260 Campus Point Drive, San Diego, CA 92121}
\date{November, 22 2010}
\maketitle

\begin{abstract}
The flow field generated by a transom-stern hullform
is a complex, broad-banded, three-dimensional phenomenon
marked by a large breaking wave. This unsteady multiphase
turbulent flow feature is difficult to study experimentally
and simulate numerically.  The results of a set of numerical
simulations, which use the Numerical Flow Analysis (NFA) code, 
of the flow around the Model 5673 transom stern  at speeds 
covering both wet- and dry-transom operating conditions are 
shown in the accompanying fluid dynamics video. The numerical predictions for 
wet-transom and dry-transom conditions are presented to demonstrate
the current state of the art in the simulation of ship
generated breaking waves.  The interested reader is referred to Drazen
et al. (2010) for a detailed and comprehensive comparison with
experiments conducted at the Naval Surface Warfare Center Carderock
Division (NSWCCD). 
\end{abstract}

\section{Computational Method}
  The Numerical Flow Analysis (NFA) code provides turnkey capabilities to
model breaking waves around a ship, including both
plunging and spilling breaking waves, the formation of
spray, and the entrainment of air. A description of NFA
and its current capabilities can be found in Dommermuth
et al. (2007); O'Shea et al. (2008); and Brucker et al.
(2010). NFA solves the Navier-Stokes equations utilizing
a Cartesian-grid formulation. The flow in the air and
water is modeled, and as a result, NFA can directly model
air entrainment and the generation of droplets. The interface
capturing of the free surface uses a second-order accurate,
volume-of-fluid technique. A cut-cell method is
used to enforce no-flux boundary conditions on the hull.
A boundary-layer model has been developed (Rottman et al. 2010), but it is not
used in these numerical simulations, and as a result, the
tangential velocities are free to slip over the hull. NFA
uses an implicit sub-grid scale (SGS) model that is built
into the treatment of the convective terms in the momentum
equations (Brucker et al. 2010). A surface representation
of the ship hull is all that is required as input
in terms of hull geometry. The numerical scheme is implemented
 in a distributed memory parallel computing environment using Fortran 90
and the  Message Passing Interface II (MPI2). Relative to methods
that use a body-fitted grid, the potential advantages of
NFA's approach are significantly simplified griding requirements
and greatly improved numerical stability due
to the highly structured grid.

\section{Domain, Grids, Boundary and Simulation Conditions}
\begin{figure}[tpb]
\centering
\includegraphics[width=5cm]{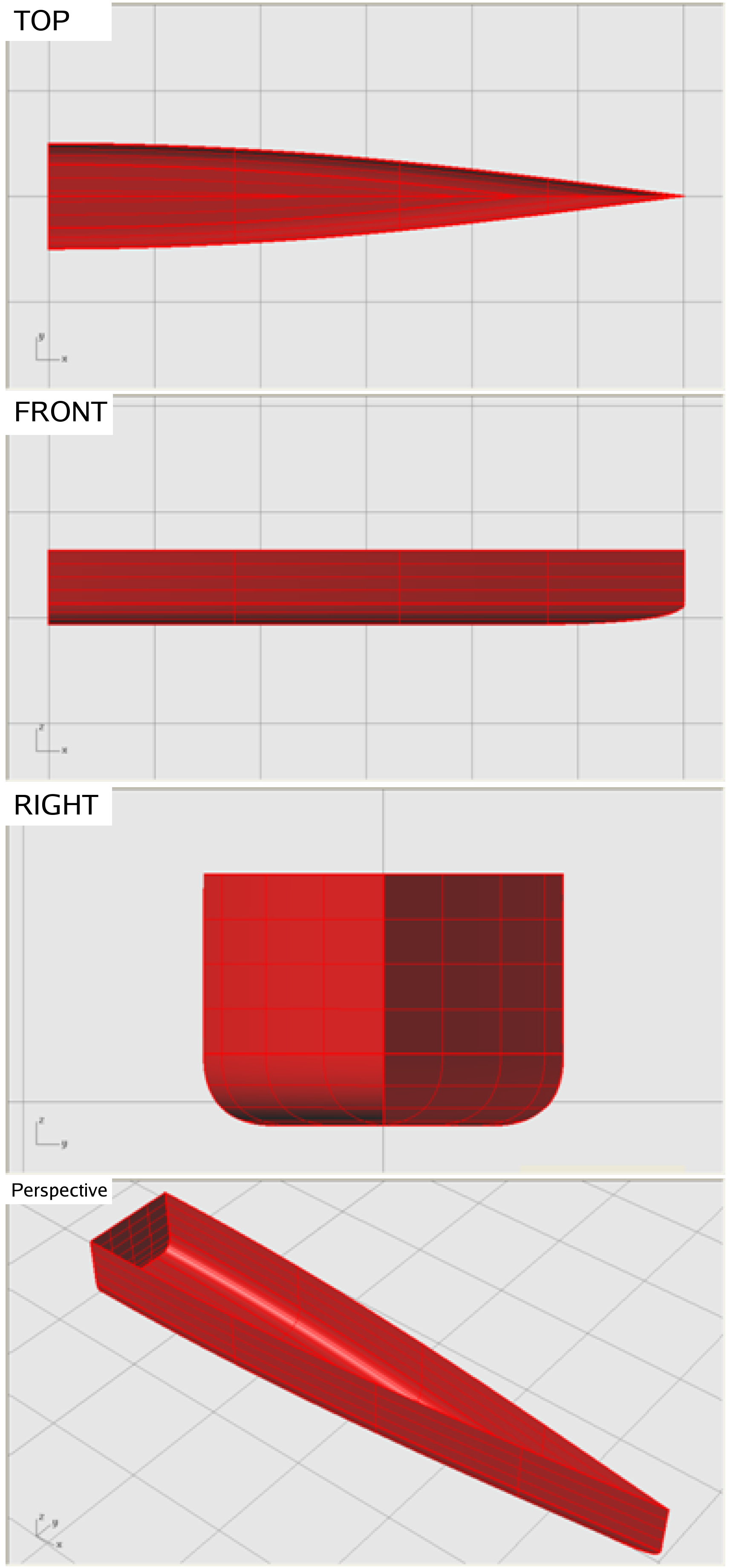}
\caption{\label{fig:1} Model 5673. Rendered at scale. Grid line
separation is 1.524 m (5 ft).}
\end{figure}

  The Model 5673 transom is shown in figure~\ref{fig:1}, and the
relevant parameters, including
the length of the model, the depth of the transom, and the weight, are given in Table~\ref{table:1}.
%

Table~\ref{table:2} provides details of the transom-stern simulations,
including the speed of the model, the Froude and Reynolds numbers, as
well as the sinkage and trim. For 3.60 m/s (7 knots), the transom is partially wet, and for 4.12
m/s (8 knots), the transom is dry. All length and velocity scales are respectively normalized by the model's length
(Lo) and speed (Uo).
The number of grid cells along the x, y, and z-axes
are respectively denoted by $N_x$; $N_y$; and $N_z$. The number
of sub domains and processors along the x, y, and
z-axes are respectively denoted by $n_i$; $n_j$ ; and $n_k$. For the
simulations shown in the fluid dynamics video, discussed herein, 
$N_x$, $N_y$, $N_z$ = $2688$, $1024$, $384$ and  
 $n_i$, $n_j$, $n_k$ = $21$, $8$, $6$.  

The width, depth, and height of the computational domains
are respectively 6.0, 1.6983, 0.66667, ship lengths
(Lo). These dimensions match the cross section of the
NSWCCD towing tank (Drazen et al. 2010). The computational domain extends
4 ship lengths behind the transom and 1 ship length
ahead of the bow. The fore perpendicular and transom
are respectively located at $x = 0$ and $x = -1$. The
z-axis is positive up with the mean waterline located at
$z = 0$. A plane of symmetry is not used on the centerline
of the hull because small-scale turbulent structures
are adversely affected.
Grid stretching is employed in all directions. Details
of the grid-stretching algorithm are provided in Dommermuth et al.
(2007).  The smallest grid spacing is 0.0005 near the hull and mean
waterline, and the largest grid spacing is $0.01$ in the far
field. The numerical simulations are slowly ramped up
to full speed. The period of adjustment is $To = 0.5$ (Dommermuth et al.
2007). Mass conservation is ensured
using the regridding algorithm that is implemented
by Dommermuth et al. (2007). Density-weighted velocity
smoothing is used every 400 time steps using a 3-
point filter ($1/4$, $1/2$, $1/4$) (Brucker et al. 2010). The nondimensional
time step is $t = 0.00025$.

The simulations are run for 30,000 time steps, corresponding
to 7.5 ship lengths, on the SGI Altix ICE at the
U.S. Army Engineer Research and Development Center
(ERDC). The data sets are so large that only time steps
20,0000 through 30,000 are saved every 40 time steps
for the purposes of post processing. The 1.06 billion cell
simulation takes about 90 hours of wall-clock time to run
30,000 time steps using 1008 processors. The wall-clock
time can be cut in half by doubling the number of processors
because NFA scales linearly. Alternatively, increasing
the number of processors to 10,000-20,000 will enable
numerical simulations of breaking ship waves and
Tsunamis with 25-50 billion grid cells within the next
year.
%
%
\begin{table}[t]
\begin{center}
\caption{\label{table:1} Table 1 Model 5673 Details.}
\vspace{0.5cm}
\begin{tabular}{|c|c|}
\hline
\hline
Length Overall, $Lo$       & 9.144 m (30 ft) \\
\hline
Waterline Length           & 9.144 m (30 ft) \\
\hline
Extreme Beam               & 1.524 m (5 ft)  \\
\hline
Bow Draft                  & Fixed           \\
\hline
Stern Draft, $T_{\it max}$ & Variable        \\
\hline
Construction               & Fiberglass      \\
\hline
Displacement               &  771.1 kg (1700 lb) \\
\hline
\hline
\end{tabular}
\end{center}
\end{table}
%
%
\begin{table}[t]
\begin{center}
\caption{\label{table:2} Calculated trim angle, 
Froude numbers based on ship length, $Fr_L$, Reynolds number based
on ship length, $Re_L$, and transom-condition.} 
\vspace{0.5cm}
\begin{tabular}{|c|c|c|c|c|c|}
\hline
\hline
 Speed & $Fr_L$ & $Re_L$ & Trim &  Transom\\
\hline
 (kts) &  & & (deg) & (m) &   Condition  \\
\hline
 7 & 0.38 & $3.29 \times 10^7$ & 0.48 & Wet \\
\hline
 8 & 0.43 & $3.77 \times 10^7$ & 0.67 & Dry\\
\hline
\hline
\end{tabular}
\end{center}
\end{table}

\section{Flow description}
The fluid dynamics video compares perspective views of laboratory
and NFA results for 3.60 m/s (7 knots) and 4.12 m/s (8
knots).  The 0.5 isosurface of the volume fractions are
shown for the NFA predictions. 
 The transom is partially wet for 3.60 m/s (7 knots) and fully dry for
4.12 m/s (8 knots). A glassy region is evident behind the transom at the 4.12 m/s (8 knots) speed. Significant
air entrainment occurs for the 3.60 m/s (7 knot) case at the transom, in the rooster-tail region, and along the
edges of the breaking stern wave. For the 4.12 m/s (8knot) case, air entrainment first occurs on the forward
face of the rooster tail and along the edges of the breaking stern wave. For both the 3.60 m/s (7 knots) and
4.12 m/s (8 knots) speeds, the measured mean profile of the free-surface elevation agrees well with instantaneous
predictions. The fluid dynamics video is available at \href{http://www.saic.com/nfa}{www.saic.com/maritime/nfa}.

\section{Video}
The full size video, mpeg4 encoded, is approximately $140Mb$
(\href{http://ecommons.library.cornell.edu/bitstream/1813/17527/2/Hand_etal_2010_APS_DFD_GalleryFluidMotion.mpg}{download}).
The web size video, mpeg4 encoded is approximately $9Mb$
(\href{http://ecommons.library.cornell.edu/bitstream/1813/17527/3/Hand_etal__2010_APS_DFD_GalleryFluidMotion_small.mpg}{download}).
\vspace{12pt}
\begin{center}
The videos are also available at \href{http://www.saic.com/nfa}{www.saic.com/maritime/nfa}.
\end{center}
\vspace{12pt}
\noindent{\bf SCENES:}\\

{\bf $Fr=0.38$, partially wetted transom:}\\
\begin{enumerate}
\item - Stern-quartering view.\\
\item - Stern view.\\
\item - Starboard view.\\
\item - Stern-quartering view compared to snapshot of experiments.\\
\item - Stern view looking down compared to snapshot of experiments.\\
\end{enumerate}

{\bf $Fr=0.43$, dry transom:}\\
\begin{enumerate}
\item - Stern-quartering view.\\
\item - Stern view.\\
\item - Starboard view.\\
\item - Stern-quartering view compared to snapshot of experiments.\\
\item - Stern view looking down compared to snapshot of experiments.\\
\end{enumerate}

\section{Acknowledgments}
We would like to acknowledge Dr. Patrick Purtell with the United States Office of Naval Research for the support through ONR grant N00014-07-C-0184.
We would like to acknowledge Dr. Thomas C. Fu and the NSWCCD for their
collaboration and continued support. 

  This work was supported in part by a grant of computer time from the
\href{http://www.hpcmo.hpc.mil}{DOD High Performance Computing
Modernization Program}. The numerical simulations have been performed on
the SGI Altix ICE-8200 at the U.S. Army Engineering Research and Development Center.

\section{References}
\noindent Brucker, K. A., O'Shea, T. T., Dommermuth, D. G., \& Adams, P.
(2010) Three-Dimensional Simulations of Deep-Water Breaking Waves.  Proceedings of the 28th
Symposium on Naval Hydrodynamics, Pasadena, California, USA, 2010.\\

\noindent Dommermuth, D.G., O'Shea, T.T., Wyatt, D.C., Ratcliffe, T., Weymouth, G.D., Hendrickson, K.L., Yue, D.K.P., Sussman, M., Adams, P., \& Valenciano, M. (2007) An application of cartesian-grid and volume-of-fluid methods to numerical ship hydrodynamics.   In the proc. of the 9th Int. Conf. on Num. Ship Hydro., Ann Arbor, MI, Aug. 5-8.\\

\noindent Drazen, D. A., Fullerton, A. M., Fu, T. C., Beale, K. L.,
O'Shea, T. T., Brucker, K. A., Wyatt, D. C., Bhushan, S., Carrica, P.
M., and Stern, F. (2010) Comparisons of model-scale experimental measurements and
computational predictions for the transom wave of a large-scale transom
model. Proceedings of the 28th Symposium on Naval Hydrodynamics, Pasadena, California,
USA. \\

\noindent O'Shea, T. T., Brucker, K. A., Dommermuth, D. G., \& Wyatt,
D. C. (2008) A numerical formulation for simulating free-surface
hydrodynamics. Proceedings of the 27th Symposium on Naval Hydrodynamics,
Seoul, Korea.\\

\noindent Rottman, J.W., Brucker, K. A., Dommermuth, D. G., \& Broutman,
D. (2010) Parameterization of the internal-wave field generated by a submarine and its turbulent wake
in a uniformly stratified fluid. Proceedings of the 28th Symposium on
Naval Hydrodynamics, Pasadena, California, USA.\\
\end{document}